# PLANCHER SOLAIRE DIRECT MIXTE À DOUBLE RÉSEAU EN HABITAT BIOCLIMATIQUE
## Conception et bilan thermique réel

*par T. DE LAROCHELAMBERT (\*)*

**Résumé.** L'article présente une nouvelle technique de Plancher Solaire Direct épais à double réseau permettant l'utilisation conjointe du chauffage solaire et d'un chauffage d'appoint. Conçue pour garantir le stockage et la diffusion de la totalité de l'énergie solaire disponible en régulant physiquement l'appoint par l'apport solaire sans gestion informatique centralisée, cette technique simple est testée et suivie dans des conditions réelles d'utilisation en habitat bioclimatique pour étudier l'influence d'une enveloppe sans inertie à grand apport solaire passif sur la productivité de l'installation solaire. Des bilans journaliers, mensuels et annuels effectués sur trois ans, complétés par des mesures en temps réel sur site, ont permis de vérifier les propriétés fonctionnelles attendues de cette technique (stockage solaire, déphasage et lissage thermique, asservissement du circuit d'appoint, économie de l'énergie d'appoint). Une analyse du fonctionnement et du bilan global à travers les concepts de productivité solaire horaire, d'énergie primaire économisée et de taux de couverture solaire corrigé est proposée pour comparer les performances énergétiques de différents types d'installations solaires.

**Abstract.** *This study presents a new Direct Solar Floor Heating technique with double heating network which allows simultaneous use of solar and supply energy. Its main purpose is to store and to diffuse the whole available solar energy while regulating supply energy by physical means without using computer controlled technology. This solar system has been tested in real user conditions inside a bioclimatic house to study the interaction of non-inertial and passive walls on the solar productivity. Daily, monthly and annual energy balances were drawn up over three years and completed by real-time measurements of several physical on-site parameters. As a result the expected properties of this technique were improved. The use of per-hour solar productivity, saved primary energy and corrected solar covering ratio is recommended to analyze the performances of this device and to allow more refined comparisons with other solar systems.*

---

(\*) Laboratoire Gestion des Risques et Environnement, Université de Haute-Alsace, CNRS EP J0082
    25 rue de Chemnitz, 68200 Mulhouse  (France)

### *Nomenclature*

| | | |
|---|---|---|
| $B$ | coefficient thermique réglementaire tenant compte des apports gratuits | $W.m^{-3}.K^{-1}$ |
| $B_{opt}$ | coefficient optique des capteurs | |
| $C$ | facteur amplificateur de correction d'ambiance | |
| $D$ | durée de fonctionnement de l'installation solaire | h |
| $DJU$ | degrés jours unifiés (base 19) | K.j |
| $E$ | éclairement solaire global | $W.m^{-2}$ |
| $EA$ | énergie d'appoint consommée | kWh |
| $EA_0$ | énergie d'appoint produite en sortie de chaudière | kWh |
| $EA_D$ | énergie d'appoint injectée dans la dalle | kWh |
| $EE$ | énergie économisée par l'installation solaire | kWh |
| $ES$ | énergie solaire captée et distribuée | kWh |
| $ES_0$ | énergie solaire globale reçue par les capteurs | kWh |
| $ES_D$ | énergie solaire injectée dans la dalle | kWh |
| $ES_{ECS}$ | énergie solaire injectée dans le ballon d'eau chaude sanitaire | kWh |
| $G$ | coefficient thermique réglementaire | $W.m^{-3}.K^{-1}$ |
| $HS_M$ | durée d'insolation mensuelle | $h.mois^{-1}$ |
| $i$ | angle d'inclinaison par rapport à l'horizontale | ° |
| $K_C$ | coefficient de transmission thermique des capteurs solaires | $W.m^{-2}.K^{-1}$ |
| $PS$ | productivité solaire | $kWh.m^{-2}$ |
| $PSH$ | productivité solaire horaire | $W.m^{-2}$ |
| $Q_v$ | débit volumique de ventilation | $m^3.h^{-1}$ |
| $R$ | ratio surface de capteurs solaires/surface de dalle solaire | |
| $R_B$ | ratio volume de stock E.C.S./surface de capteurs solaires | m |
| $S_C$ | surface de capteurs solaires | m² |
| $S_D$ | surface de plancher solaire direct | m² |
| $S_H$ | surface habitable | m² |
| $S_{HP}$ | surface habitable pondérée | m² |
| $T_{ADD}$ | température de départ de circuit d'appoint dalle | °C |
| $T_{ARD}$ | température de retour de circuit d'appoint dalle | °C |
| $T_{BD}$ | température de départ du circuit solaire E.C.S. | °C |
| $T_{BR}$ | température de retour du circuit solaire E.C.S. | °C |
| $T_{CB}$ | température de consigne de surchauffe de ballon E.C.S | °C |
| $T_{CD}$ | température de consigne de surchauffe de dalle solaire | °C |
| $T_{CI}$ | température de consigne intérieure | °C |
| $T_{CP}$ | température de consigne de protection de dalle solaire | °C |
| $T_E$ | température extérieure | °C |
| $T_I$ | température intérieure | °C |
| $T_{SDD}$ | température de départ de circuit solaire dalle | °C |
| $T_{SRD}$ | température de retour de circuit solaire dalle | °C |
| $V_{ECS}$ | volume d'eau chaude sanitaire consommé | m³ |
| $V_H$ | volume habitable | m³ |

*Symboles grecs*

| | | |
|---|---|---|
| $\Delta t_{SA}$ | temps de transit du réseau solaire au réseau d'appoint | s |
| $\Delta t_{SO}$ | temps de transit du réseau solaire à la surface de dalle | s |
| $\tau_{MB}$ | taux de couverture solaire mensuel brut | % |
| $\tau_{MC}$ | taux de couverture solaire mensuel corrigé | % |
| $\tau_{AB}$ | taux de couverture solaire annuel brut | % |
| $\tau_{AC}$ | taux de couverture solaire annuel corrigé | % |
| $\eta$ | rendement de l'installation solaire | % |
| $\eta_M$ | rendement solaire mensuel | % |

*Indices*

| | |
|---|---|
| $A$ | annuel |
| $J$ | journalier |
| $M$ | mensuel |

## 1. INTRODUCTION

La technique du Plancher Solaire Direct (P.S.D.), conçue et mise au point par l'École Supérieure d'Ingénieurs de Marseille (E.S.I.M.) il y a une quinzaine d'années [1], a permis une diffusion plus importante du chauffage solaire dans l'habitat individuel grâce à la réduction des coûts d'investissement et la simplicité de mise en oeuvre qu'elle entraîne par rapport aux systèmes conventionnels de chauffage solaire [2].

Dans ces derniers en effet, l'énergie absorbée par les capteurs solaires est stockée dans de grands réservoirs d'eau par l'intermédiaire d'échangeurs, l'eau ainsi chauffée étant distribuée à basse température dans les émetteurs de chaleur (radiateurs et planchers chauffants). Les inconvénients de ces systèmes sont principalement:
- une perte importante de rendement due à la présence d'échangeurs;
- les déperditions thermiques importantes des ballons de stockage, généralement extérieurs au volume habitable;
- un coût très élevé dû à l'échangeur (titane ou cuivre) et aux ballons
- une multiplication des circulateurs, des régulations, des vannes motorisées et donc des risques de pannes, de fuites ou de dysfonctionnement.

Le P.S.D. dans son principe non seulement pallie ces inconvénients mais offre de surcroît des atouts décisifs que nous rappelons brièvement [3]:
- stockage thermique de l'énergie solaire dans le plancher en béton, intérieur au volume habitable;
- inertie thermique importante;
- absence d'échangeur;
- régulation simplifiée, généralement réduite à un thermostat différentiel contrôlant un circulateur unique;
- avantages des planchers chauffants: confort basse température, uniformité de température de l'air ambiant, suppression possible des radiateurs muraux;
- économie d'investissement importante.

Depuis la mise au point de la technique de chauffage par P.S.D., plusieurs centaines d'installations à P.S.D. ont été réalisées, presqu'exclusivement en France, essentiellement dans le secteur de l'habitat individuel et quelques unes dans le petit collectif ou le secteur hospitalier. Divers suivis et campagnes de mesures ont permis de vérifier le bon comportement des dalles solaires et leur apport positif dans les bilans énergétiques, annonçant des « taux de couverture solaire » généralement compris entre 30 à 50% [4][5][6][7].

Cependant, un des obstacles majeurs à une diffusion plus générale de la technique P.S.D. est la nécessité d'investir dans un chauffage d'appoint performant, devant couvrir la totalité de la puissance de chauffe du bâtiment pendant d'éventuelles longues périodes sans ensoleillement en hiver, sans pouvoir disposer du chauffage par le sol. La solution consistait jusqu'à présent à doubler le P.S.D. d'un système complet de chauffage classique par radiateurs ou par poêles à bois, fuel ou charbon, ou par cheminées à insert. Seule la perspective d'importantes économies de chauffage engendrées par le P.S.D. pouvait compenser le surinvestissement dans le chauffage d'appoint et la perte de confort du plancher chauffant en l'absence de soleil [8].

## 2. COUPLAGES SOLAIRE-APPOINT DANS LE PLANCHER CHAUFFANT

Divers travaux récents ont tenté d'apporter une solution économique au problème, en cherchant à utiliser les dalles de P.S.D. comme planchers chauffants utilisés simultanément par les circuits solaires et par les circuits de chauffage d'appoint.

### 2.1. P.S.D. mince à appoint intégré

Cette solution a été étudiée et mise en oeuvre par P.PAPILLON [9][10] et G.ACHARD et al. [11] et la société T2I, fabricant de panneaux et de systèmes solaires en Savoie (France). Elle consiste pour l'essentiel en un couplage hydraulique et thermique des circuits d'appoint et solaire dans un seul et même réseau de tubes chauffants noyés dans le plancher solaire en béton, dont l'épaisseur est diminuée de moitié par rapport aux P.S.D. classiques (15 cm au lieu de 30 cm). Les auteurs ont simulé numériquement le fonctionnement d'un tel système, basé sur une gestion centralisée de l'ensemble par microprocesseur, avec identification de paramètres *in-situ* et auto-adaptation, puis vérifié leur modèle sur un P.S.D. expérimental. Il apparaît que l'épaisseur de la dalle est un paramètre fondamental du comportement thermique du P.S.D. et de son interaction avec l'enveloppe de l'habitation.

Ainsi une dalle épaisse (supérieure à 20 cm) offre une inertie suffisamment importante pour rendre négligeable l'effet du type d'isolation (intérieure ou extérieure) des murs de l'habitation. Les auteurs soulignent que l'apport thermique du chauffage solaire "actif" par un P.S.D. est d'autant plus faible que les apports solaires "passifs" à travers les vitrages sud sont plus élevés; ils concluent à l'incompatibilité d'une enveloppe à isolation intérieure (faible inertie) et à grands apports solaires passifs vis à vis des performances d'un P.S.D., et préfèrent une solution dalle mince (15 cm) dans un habitat classique (forte inertie des murs pour compenser la perte d'inertie de la dalle; peu d'apports solaires passifs). La baisse de productivité solaire ainsi engendrée est alors compensée par une diminution du coût d'investissement et par une augmentation de la production solaire d'eau chaude sanitaire (E.C.S.).

Il est clair que cette solution présente un intérêt majeur en habitat collectif où l'emploi de dalles minces s'impose, et devrait permettre au chauffage solaire de pénétrer plus efficacement ce secteur où d'énormes économies d'énergie pourraient ainsi être engendrées.

En revanche l'opposition entre performances du P.S.D. et performances thermiques de l'habitat ne nous apparaît pas justifiée si l'on considère l'ensemble {chauffage solaire + chauffage d'appoint + enveloppe bioclimatique + E.C.S.} conçu autour d'un plancher chauffant à double nappe de chauffage tel que nous le présentons ci-dessous.

## 2.2. P.S.D. mixte à double réseau

Nous avons étudié dès 1990 la faisabilité technico-économique d'un système de chauffage global, basé sur le concept de *couplage thermique asservi du réseau de chauffage d'appoint par le réseau de chauffage solaire dans un seul et même plancher chauffant, avec découplage hydraulique total*. Cette étude, basée sur la méthode E.S.I.M. [1][2] et complétée par une modélisation monodimensionnelle, a été menée dans le cadre global d'un habitat moderne performant de type bioclimatique de manière à intégrer les deux paramètres thermiques fondamentaux caractérisant la réponse de l'enveloppe, à savoir les apports solaires passifs gratuits (que l'on cherche à maximiser) et l'inertie des parois extérieures (que l'on rend minimale). Elle s'est inscrite dans une action incitative régionale basée sur un programme de dix-huit installations à P.S.D. réunissant l'Agence de l'Environnement et de la Maîtrise de l'Énergie (A.D.E.M.E.), la Région Alsace et l'association Alter Alsace Énergies [12].

*Elle s'est fixée comme objectifs l'évaluation expérimentale précise sur une longue période (plusieurs années) de la réponse thermique d'une dalle solaire à double nappe en habitat bioclimatique à toutes les séquences climatiques possibles, et la détermination comparée des performances énergétiques mensuelles et annuelles de ce système obtenue à partir des mesures expérimentales d'une part, et par la méthode E.S.I.M. d'autre part. Les objectifs ultérieurs sont la mise au point d'une simulation numérique globale du système et l'affinement des critères de dimensionnement et des bilans prévisionnels mensuels et annuels.*

Ce travail mené en collaboration avec l'Atelier Architecture et Soleil de Strasbourg (67) et la société Éco-Chauffage de Ribeauvillé (68) a conduit à la construction en 1991 d'une maison prototype bioclimatique à ossature bois dans la plaine d'Alsace, à la latitude de MULHOUSE, sur un site parfaitement dégagé près du Rhin, exposé à tous les vents et aux brouillards fréquents dans cette région *(figure 1)*. On peut le considérer comme un cas représentatif de conditions climatiques défavorables. Une instrumentation résidente complétée par des relevés continus par acquisition informatique nous permet de dresser un *bilan énergétique expérimental* précis de l'installation à P.S.D. mixte et de l'habitat, et de vérifier le comportement thermique du P.S.D. mixte dans les situations les plus diverses en fonctionnement réel, la maison étant habitée depuis août 1991 [13].

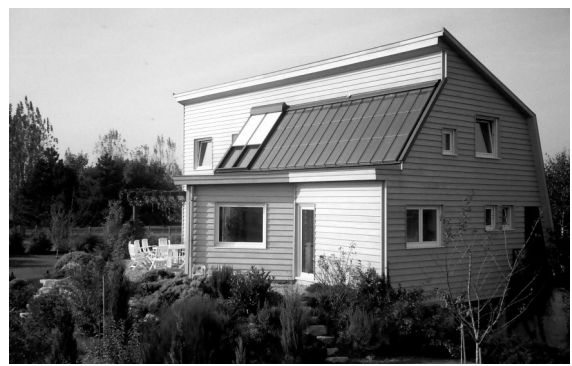

*Fig. 1. – Vue de l'habitation bioclimatique à PSD mixte près de Mulhouse (France)*
Fig.1. – View of bioclimatic house with mixed direct solar floor near Mulhouse (France)

## 3. CONCEPTION DU SYSTÈME SOLAIRE À P.S.D. MIXTE

### 3.1. Enveloppe thermique de l'habitation

La conception bioclimatique de l'habitation repose sur les caractéristiques suivantes:
- orientation sur les points cardinaux avec vitrages principaux au sud, dont une grande verrière intégrée, occultable extérieurement;
- toiture entièrement au nord jusqu'au sol, en tuiles épaisses de terre cuite, avec 30 cm de laine de verre, sans grenier;
- une seule ouverture vitrée à l'ouest (pluies et vents dominants);
- protection des entrées à l'est et vitrages plus nombreux;
- cave (garage, cellier et chaufferie-buanderie-séchoir) sous toute la maison;
- conception interne: communication ouverte entre rez-de-chaussée (R.d.C.) et étage par escalier ouvert sous la verrière et mezzanines, dans le but d'uniformiser la température, de transmettre le rayonnement du P.S.D. à l'étage et l'éclairement de la verrière au R.d.C.;
- choix des matériaux: l'ossature et les murs extérieurs sont en bois, avec double isolation laine de roche, laine de verre; le plancher d'étage est en bois avec isolation phonique. Le choix du bois permet une forte isolation thermique et une très faible inertie thermique. Cette faible inertie est également recherchée pour les parois intérieures en placostil doublé avec laine de verre. Les huisseries extérieures sont en PVC, et tous les vitrages sont doubles à faible pouvoir émissif.

L'ensemble de ces choix conduit aux caractéristiques générales suivantes:
- surface habitable $S_H$ = 132 m² ($S_{HP}$ = 187 m²)
- volume habitable $V_H$ = 330 m³
- coefficients thermiques calculés $G$ = 0,642 W.m$^{-3}$.K$^{-1}$ , $B$ = 0,375 W.m$^{-3}$.K$^{-1}$
- ventilation mécanique contrôlée simple flux $Q_v$ = 100 m³.h$^{-1}$.

### 3.2. Installation solaire à P.S.D. mixte

Le Plancher Solaire Direct Mixte est constitué d'une dalle de béton épaisse (26 cm) coulée sur toute la surface de rez-de-chaussée sur hourdis en béton armé Fricker, fortement isolée sur sa face inférieure côté cave par 23 cm de polystyrène et fibralith, comportant un double réseau de tubes de chauffage *(figure 2)*:
- *le réseau solaire* directement relié aux capteurs solaires: de forte densité (pas de 20 cm entre tubes), il est noyé à environ 3 cm du fond de la dalle de manière à charger thermiquement tout le volume de la dalle et à imposer un gradient de température vers le haut. Cette position basse assure au P.S.D. ses *fonctions de stockage et de déphasage* indispensables à une bonne productivité solaire et au confort thermique. Un thermostat électronique différentiel commande le circulateur du réseau en tout-ou-rien;
- *le réseau d'appoint* relié à la chaudière (le gaz naturel a été choisi pour son faible coût, sa souplesse de régulation et son entretien réduit) par le biais d'une vanne motorisée trois voies assurant le bouclage du réseau et sa liaison à une bouteille de mélange chauffée à 65°C par la chaudière (qui alimente par ailleurs les petits radiateurs d'étage). Ce réseau est noyé *en surface de dalle*, à environ 7 cm sous le carrelage de manière à répondre rapidement aux variations des besoins de chauffe par une régulation classique de plancher chauffant par vanne mélangeuse à sondes extérieure, d'ambiance et de départ de circuit, tout en assurant une diffusion de chaleur suffisante pour homogénéiser la température superficielle de la dalle. Ce réseau est 1,5 fois moins dense que le réseau solaire (30cm entre tubes).

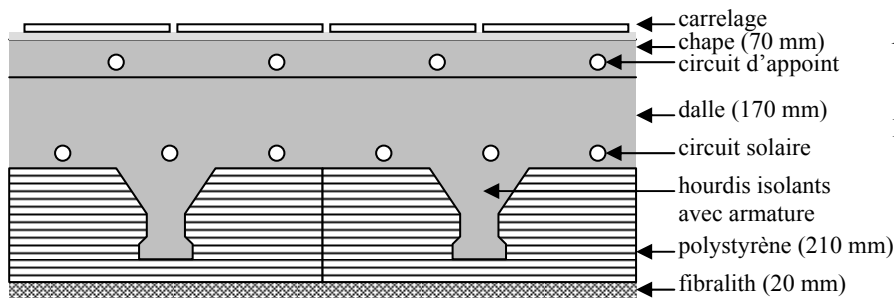

*Fig. 2. – Schéma simplifié du PSD mixte (coupe transversale)*
Fig.2. – Simplified plan of mixed direct solar floor cross-section

Le principe physique autour duquel est conçu le P.S.D. mixte repose sur *l'interaction thermique unidirectionnelle* des deux réseaux: l'apport solaire en fond de dalle crée une densité de flux de chaleur ascendante dont la période caractéristique est d'une journée, qui assure l'asservissement thermique de l'appoint dans le réseau de surface par le biais de deux phénomènes, à savoir le *temps de transit caractéristique des variations thermiques entre le réseau solaire et le réseau d'appoint* $\Delta t_{SA}$ = 3,9 h (calculé selon un modèle monodimensionnel de diffusion en régime établi du flux solaire injecté par journées ensoleillées successives dans une dalle isolée sur une face, avec coefficients d'échange constants) suffisamment faible pour permettre une influence rapide après une période sans soleil, et la *température moyenne du P.S.D.* qui dépend des apports solaires antérieurs. Le rôle du réseau d'appoint se limite alors approximativement aux 10 cm superficiels de la dalle car les variations de sa température, de périodes plus courtes que la journée, sont plus fortement amorties et se propagent plus rapidement.

C'est à ce niveau qu'intervient la conception bioclimatique de l'habitation: la régulation de l'appoint dans le réseau de surface est classiquement déterminée selon la loi de chauffe (cf. § 3.3) par la température de son départ (élevée si la dalle a été chauffée antérieurement par le circuit solaire), la température extérieure et surtout la température intérieure de l'habitation. La très faible inertie thermique des parois associée aux grandes ouvertures Sud et Est permet une élévation suffisamment rapide de la température ambiante pour que la régulation du circuit d'appoint réagisse très rapidement aux apports solaires passifs grâce à sa fonction de correction d'ambiance qui abaisse la température de départ du réseau d'appoint de la quantité $C(T_{CI} - T_I)$, réalisant ainsi une véritable *accélération de l'effacement de l'appoint devant le solaire*. Cette action est renforcée par l'absorption de chaleur par la surface du P.S.D. directement éclairée par le soleil à travers les vitrages.

L'effet prévu est de refermer rapidement la vanne de bouclage du circuit d'appoint pour éviter le maintien du chauffage de la partie supérieure de la dalle par ce circuit alors que le réseau solaire doit charger celle-ci. Comme l'enveloppe de l'habitation est très isolée, peu inerte et fortement passive, il est vérifié qu'il ne résulte pas de baisse de température ambiante de cet effacement de l'appoint en surface de dalle, les dix centimètres superficiels restituant la chaleur stockée suffisamment lentement à l'air ambiant.

Le *choix de l'épaisseur totale de la dalle* est fait en fonction du temps de stockage prévu (deux à trois jours en intersaison), du niveau moyen de température de dalle souhaité pour permettre l'autonomie journalière en hiver par période d'ensoleillement continu (température ambiante 19°C compte tenu des apports passifs), du déphasage voulu entre l'éclairement maximal et la restitution de chaleur en surface. Une trop grande épaisseur (supérieure à 30 cm) conduit à une température moyenne trop basse, à un $\Delta t_{SA}$ et à un temps de transit $\Delta t_{SO}$ du flux de chaleur solaire jusqu'à la surface trop grands (supérieurs à 5 h et 7,5 h respectivement). Inversement, une trop faible épaisseur (inférieure à 20 cm) entraîne un risque de réchauffement du circuit solaire par le circuit d'appoint et de surchauffe de la dalle en intersaison, un inconfort dû aux amplitudes thermiques trop fortes en surface de dalle, et conduit à une diminution du stockage solaire, ayant pour conséquence une demande d'appoint plus grande pour assurer les besoins.

L'épaisseur de dalle retenue de 26 cm (avec carrelage), la distance de 16 cm environ entre les deux réseaux, et de 7 cm entre le réseau d'appoint et la surface de dalle constituent une bonne solution de compromis compatible avec le caractère bioclimatique de l'enveloppe ($\Delta t_{SO}$ = 6h, $\Delta t_{SA}$ = 3,9 h, temps de déstockage de 3 jours en octobre), une dalle de compression de 17 cm étant coulée sur la structure en hourdis isolants avec armature. Ces chiffres peuvent évidemment varier de ± 1 cm lors de la mise en oeuvre du P.S.D.. La distance de 7 cm entre circuit d'appoint et surface peut être légèrement diminuée jusqu'à 5 cm, mais avec des risques d'inconfort (le réseau étant lâche, la température au droit des tubes est alors plus élevée ainsi que ses variations spatiales en surface) et de retard de l'action du circuit solaire sur le circuit gaz de près d'une demi-heure.

*L'objectif premier du P.S.D. mixte est donc de rendre compatible le chauffage solaire et la structure bioclimatique d'un bâtiment*. Cette technique offre en outre l'avantage de la simplicité dans la mise en oeuvre et dans la gestion simultanée de l'appoint et du chauffage solaire. En effet, toutes les régulations électroniques sont classiques et fiables cf. § 3.3). D'autre part, *l'indépendance hydraulique totale des deux réseaux* évite tout risque de fuite thermique de vannes de mélange, préjudiciable au rendement solaire, et permet une maintenance séparée des deux chauffages.

En outre, la chaudière peut également assurer le chauffage complémentaire de l'étage par petits radiateurs à robinets thermostatiques qui répondent très rapidement aux apports solaires passifs.

La fonction *production d'E.C.S.* est assurée simultanément par un circuit solaire, parallèlement au circuit P.S.D., avec circulateur et thermostat différentiel indépendant, et par la même chaudière d'appoint à travers un double échangeur à l'intérieur d'un ballon fortement calorifugé de 500 dm$^3$. Le circuit solaire chauffe tout le stock par l'échangeur du bas à 100% en été et le préchauffe tout le reste du temps (au moins en intersaison). L'appoint ne chauffe la moitié supérieure du stock que si sa température descend au-dessous de 45°C, ce qui provoque le basculement d'une vanne de zone permettant à la chaudière de chauffer en priorité l'eau chaude sanitaire; cette opération est rendue possible par l'inertie suffisante de la dalle et l'excellente isolation des murs sans que l'on ressente l'absence de chauffage durant cette période, même sans apport solaire.

La configuration de l'ensemble de l'installation schématisée en *figure 3* appelle quelques remarques supplémentaires:

- le circuit solaire du P.S.D. mixte est automatiquement coupé dès que la température de fond de dalle près du tube retour $T_{SRD}$ dépasse une valeur de consigne $T_{CD}$ fixée par aquastat réglable (la plage de 21°C ≤ $T_{CD}$ ≤ 23°C convient). *Ceci évite toute surchauffe de dalle solaire en intersaison* et permet de transférer alors intégralement l'énergie solaire captée au ballon d'E.C.S., ce qui augmente le rendement *global* de l'installation [9]; un second aquastat en série avec le précédent n'autorise le redémarrage du circuit solaire P.S.D. que si la température du fluide solaire en sortie des capteurs est inférieure à une température maximale de protection $T_{CP}$ de l'ordre de 45 °C pour éviter de l'injecter trop chaud dans le circuit refroidi de la dalle (conformité à la réglementation, D.T.U. 65.8 [14]);
- lorsque l'énergie solaire absorbée par l'installation risque de dépasser les besoins d'E.C.S. en été, une soupape de sécurité permet un soutirage automatique d'E.C.S., récupéré ou évacué à l'égout selon les besoins, dès que la température de l'eau en haut de ballon d'E.C.S. dépasse une température de consigne $T_{CB}$ = 95°C. De la sorte, le fonctionnement et la sécurité de l'installation sont garantis même en l'absence des propriétaires;
- la mise en parallèle des circuits solaires P.S.D. et E.C.S. avec circulateurs, thermostats différentiels et vannes d'isolement indépendants offre une grande simplicité de réglage et de maintenance;
- l'investissement global dans les circuits d'appoint est réduit (chaudière de petite puissance à ventouse sans préparation d'appoint séparé).

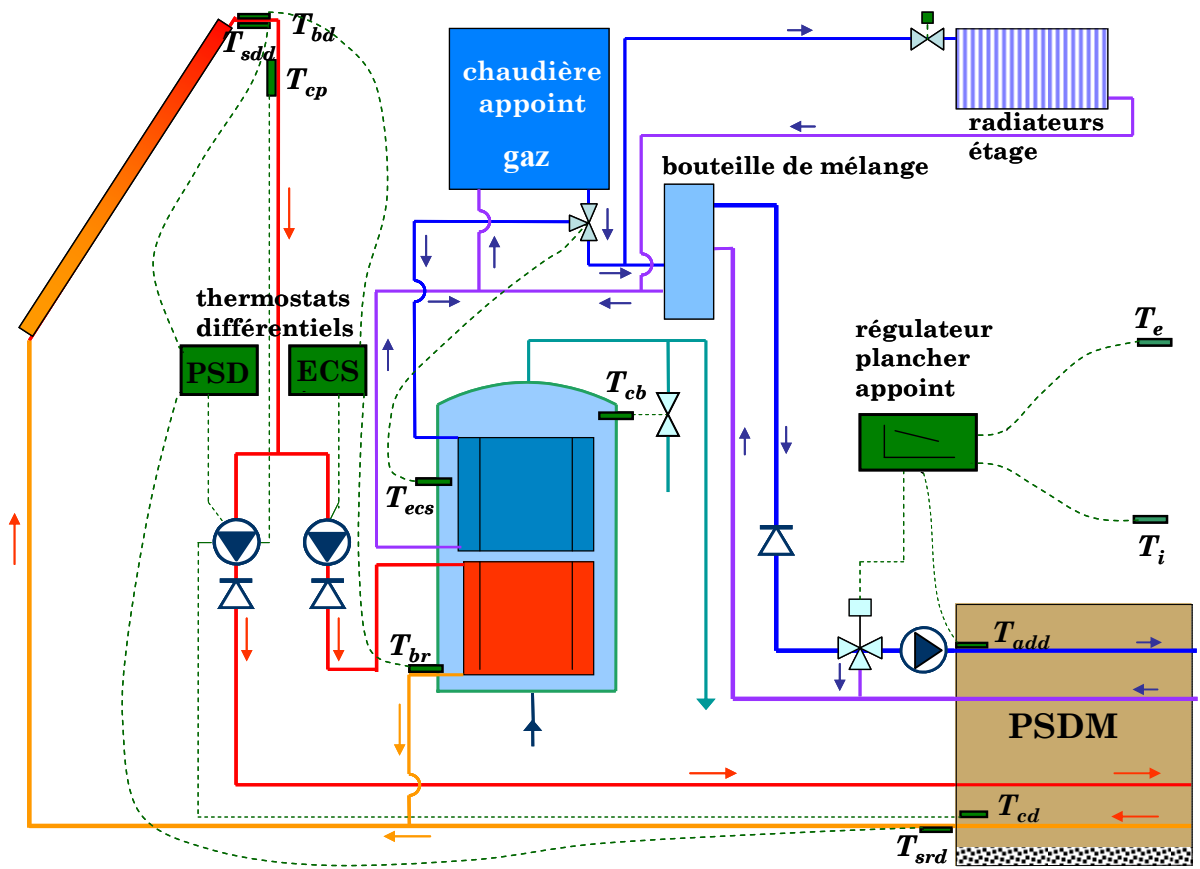

*Fig. 3. – Schéma hydraulique de l'installation complète de chauffage et d'ECS*
Fig.3. – Hydraulic flowsheet of mixed heating and SHW plant

### 3.3. Dimensionnement de l'installation et bilan énergétique prévisionnel

L'utilisation de la méthode de calcul de l'E.S.I.M. et des données de la station de Mulhouse de la Météorologie Nationale a conduit à une première approche classique du dimensionnement par bilans mensuels moyens, largement utilisée dans les bureaux d'étude, mais dont la simplicité et les hypothèses de base amènent à sous-estimer les pertes inférieures de dalle et les couplages avec l'E.C.S. et l'enveloppe du bâtiment. Elle fournit cependant une valeur indicative assez représentative de la productivité annuelle globale et du taux de couverture solaires dont nous ferons plus loin une étude critique.

La disposition du P.S.D. mixte sur plancher Fricker n'étant pas prise en compte dans la méthode E.S.I.M., divers dimensionnements ont été effectués pour déterminer une fourchette de taux de couverture mensuels et annuels amenant à une autonomie quasi totale d'avril à octobre inclus, en faisant varier les épaisseurs de dalle et d'isolant, la surface des capteurs et leur inclinaison *i*.

La configuration retenue est présentée ci-dessous:
- *installation solaire:*
  - capteurs solaires: $S_C = 17$ m$^2$; $i = 58°$; $B_{opt} = 0,68$; $K_C = 4,2$ W.m$^{-2}$.K$^{-1}$; $R = 0,191$; type sélectif intégré en façade;
  - circuit P.S.D.: $S_D = 89$ m$^2$; réseau de tube polyéthylène réticulé de pas 20 cm sur toute la surface (pas de zonage Nord/Sud); thermostat électronique différentiel tout-ou-rien avec sondes à résistance métallique;
  - circuit E.C.S. $V_{ECS} = 0,5$ m$^3$; thermostat électronique différentiel, identique au précédent;
- *chauffage d'appoint:*
  - chaudière murale à ventouse au gaz naturel de 10 kW (modèle 18 kW installé) sans préparateur d'E.C.S.;
  - réseau de surface P.S.D. en polyéthylène réticulé de pas 30 cm sur toute la surface; puissance 7000 W; régulation de plancher chauffant pente 0,4; calage jour 19°C (nuit 18°C); correction d'ambiance C = 9;
  - radiateurs d'étage à robinets thermostatiques: puissance totale 3000 W; radiateur buanderie: 2000 W;
- *particularités:*
  - aucun autre appoint n'est utilisé (poêle, cheminée, radiateur électrique, etc.);
  - la fonction séchage de linge est assurée en hiver par le radiateur de buanderie sous étendoir;
  - la fonction lavage de vaisselle est assurée uniquement par l'eau chaude du ballon d'E.C.S. (pas d'électricité)
  - le lave-linge utilise l'E.C.S. du ballon par mélangeur (économie quasi totale d'électricité).
- *site météorologique*: les données moyennes de 1951-1970 de la station de Mulhouse sont résumées dans le *tableau I*. Elles seront comparées aux données de 1992 à 1994 et aux mesures effectuées sur la maison.
- *bilan énergétique global:*
  - chauffage : 9980 kWh dont 4160 à 4300 kWh solaires
  - E.C.S. : 2070 kWh dont 710 à 720 kWh solaires
  - total : 12050 kWh dont 4880 à 5020 kWh solaires.
  - taux de couverture solaire annuel : 40,5 à 41,7% ; productivité solaire annuelle : 287 à 295 kWh.m$^{-2}$.an$^{-1}$.

| TABLEAU I – TABLE I *Données météorologiques (1951-1970)* - Meteorological data (1951-1970) | | | | | | | | | | | | | |
|---|---|---|---|---|---|---|---|---|---|---|---|---|---|
| Mois | JAN | FEV | MAR | AVR | MAI | JUN | JUL | AOU | SEP | OCT | NOV | DEC | ANNÉE |
| DJU(K.jour) | 536,3 | 462,0 | 378,2 | 285,0 | 151,9 | 45,0 | 0 | 0 | 90,0 | 248,0 | 402,0 | 499,1 | 3097,5 |
| $HS_M$(h) | 60 | 90 | 155 | 173 | 220 | 222 | 240 | 210 | 177 | 140 | 65 | 50 | 1802 |
| $T_E$ (K) | 1,7 | 2,5 | 6,8 | 9,5 | 14,1 | 17,5 | 20,0 | 19,4 | 16,0 | 11,0 | 5,6 | 2,9 | 10,6 |

### 3.4. Métrologie

Le suivi énergétique de l'habitation a été réalisé à partir des données suivantes:
- comptage quotidien de l'énergie solaire distribuée dans l'installation, de la durée de fonctionnement de l'installation solaire, de la consommation de gaz par la chaudière seule, de l'énergie d'appoint gaz distribuée en sortie de chaudière et dans le réseau de surface P.S.D., de l'eau chaude sanitaire consommée;

- acquisition permanente par centrale informatique autonome sur batterie de l'éclairement solaire global hémisphérique par pyranomètre dans le plan des capteurs; de la température extérieure à 1,5m sous abri; de la température intérieure au centre de l'habitation; des températures départ-retour des réseaux solaire et appoint dans le P.S.D.; des températures départ-retour du circuit solaire E.C.S.; et du débit volumique total du fluide caloporteur des capteurs solaires;
- relevés de la Météorologie Nationale des stations les plus proches (Mulhouse et Colmar).

## 4. BILANS ÉNERGÉTIQUES RÉELS DE L'INSTALLATION SOLAIRE

Les données recueillies permettent d'établir des bilans énergétiques quotidiens, mensuels et annuels précis du *système solaire* que constitue l'habitation bioclimatique, son installation de chauffage et d'E.C.S. mixte solaire-appoint gaz en conditions réelles d'utilisation.

Nous étudions dans un premier temps les principaux paramètres permettant un diagnostic clair du comportement global du système, pour lequel nous proposons des critères d'analyse comparative caractérisant le fonctionnement de l'installation et son potentiel énergétique, et nous exposons les bilans réels utiles que l'on peut en déduire, pour les comparer aux bilans prévisionnels classiques présentés précédemment (*cf. schéma méthodologique*).

Dans un second temps, une analyse plus fine du fonctionnement du P.S.D. mixte et de son couplage avec l'enveloppe et la production d'E.C.S. dans diverses séquences climatiques journalières caractéristiques permet d'éclairer ces bilans globaux (voir paragraphe 5).

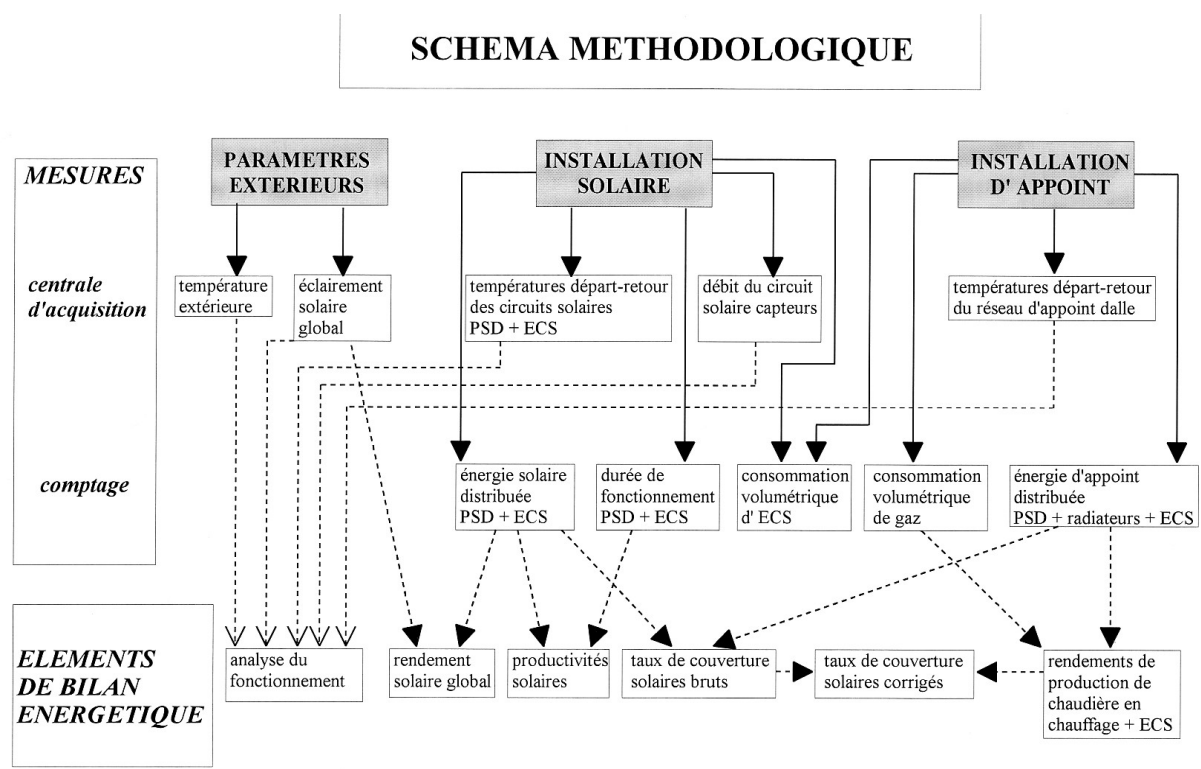

### 4.1. Paramètres d'analyse énergétique de l'installation solaire

L'analyse du fonctionnement réel de l'installation solaire nous a amenés à utiliser plusieurs paramètres d'analyse énergétique et d'en proposer de nouveaux, afin de cerner au mieux les propriétés fonctionnelles des systèmes solaires basse température et permettre leur comparaison dans des contextes de couplages variés avec l'appoint et l'enveloppe.

À partir des grandeurs globales mensuelles et annuelles mesurées par relevé quotidien ($ES_M$, $ES_A$, $EA_D$, $EA_M$, $EA_A$, $D_M$) et de $EA_0$, nous déterminons les paramètres d'analyse énergétique globaux suivants :

- *énergies économisées mensuelles et annuelles $EE_M$ et $EE_A$* : c'est l'énergie primaire d'appoint que la chaudière aurait dû consommer en plus en l'absence de l'installation solaire ; elle est obtenue par addition des énergies solaires distribuées au P.S.D. et au ballon d'E.C.S. divisées par les rendements de distribution et de production du système d'appoint dans les fonctions chauffage et production d'E.C.S. ;

- *taux de couverture solaire mensuels et annuels bruts courants $\tau_{MB}$ et $\tau_{AB}$* : ils sont définis par les relations suivantes $\tau_{MB} = 100\ ES_M/(ES_M + EA_M)$ et $\tau_{AB} = 100\ ES_A/(ES_A+EA_A)$ et ne représentent que les rapports de l'énergie solaire effectivement captée et distribuée à l'énergie totale primaire utilisée dans l'habitation ;

- *taux de couverture solaire mensuels et annuels corrigés $\tau_{MC}$ et $\tau_{AC}$* : ils donnent la part réellement couverte par l'énergie solaire *active* compte tenu de tous les rendements de production et donc plus proches de la réalité et de l'économie solaire ; nous les définirons par $\tau_{MC} = 100\ EE_M/(EE_M+EA_M)$ et $\tau_{AC} = 100\ EE_A/(EE_A+EA_A)$ ;

- *productivités solaires journalières, mensuelles et annuelles $PS_J$, $PS_M$ et $PS_A$* : classiquement utilisées en ingénierie solaire, elles mesurent l'énergie produite par mètre carré de capteur solaire installé pour une journée, un mois et une année; ces grandeurs dépendent du type de production (chauffage / E.C.S.), de la conception de l'installation (échangeurs, P.S.D., ratios $R$ et $R_B$, $i$) et surtout des données locales du site (latitude, altitude, climat) ;

- *productivité solaire horaire PSH* : nous utiliserons ce concept pour mieux définir la productivité réelle de l'installation *quand elle fonctionne*. Nous la définirons par $PSH_J = ES_J/(S_C.D_J)$ lorsque nous la calculerons sur une journée et par $PSH_M = ES_M/(S_C.D_M)$ sur un mois de fonctionnement. Nous verrons par la suite que *c'est un paramètre représentatif de l'efficacité de fonctionnement du système*, permettant la mesure de l'influence de l'appoint et de l'impact du choix de l'inclinaison $i$ à la conception. Il représente la *puissance effective moyenne* de l'installation en production sur une journée, un mois (voire un an) ;

- *rendements solaires journaliers et mensuels $\eta_J$ et $\eta_M$* : également classiques, ils mesurent le rapport de l'énergie réellement produite par l'installation solaire à l'énergie solaire globale incidente sur les capteurs. Ce sont des paramètres représentatifs du type de fonctionnement de l'installation (chauffage et/ou E.C.S.) car très sensibles aux niveaux de températures de stock et de température extérieure.

Le concept d'*énergie économisée* repose sur la substitution de l'énergie solaire à l'énergie d'appoint: c'est l'énergie primaire qu'aurait dû consommer la chaudière en l'absence de système solaire pour fournir la même énergie. Le rendement de production de la chaudière est pris en compte dans ses deux fonctions chauffage et E.C.S.; pour les chaudières de type mural à ventouse, il est généralement pris égal à 75%, mais les mesures effectuées sur site donnent des valeurs inférieures: 68,2% en chauffage seul; 51,3% en production d'E.C.S. seule; 62,5% en mode chauffage + E.C.S. (ces chiffres sont mesurés à ± 7% près). Les rendements de distribution et de stockage sont sensiblement identiques pour les circuits solaires et d'appoint, tant en chauffage qu'en E.C.S.. On peut donc estimer ici l'énergie économisée par la relation $EE = (ES_D/0{,}682) + (ES_{ECS}/0{,}513)$.

Le concept de *taux de couverture solaire corrigé* est de ce fait beaucoup plus proche de la réalité énergétique puisqu'il englobe à la fois tous ces rendements et tous les types de consommation habituels dans une maison, y compris ceux liés à la vaisselle, au séchage et au lavage de linge, au renouvellement d'air. Il tient compte d'autre part des exigences modernes de confort, en particulier la quasi uniformité de température entre étage et R.d.C.. Il est facilement mesurable sur site, contrairement au taux de couverture théorique classique, rapport entre l'énergie solaire utile et les besoins totaux hors rendements, dont il est assez proche.

### 4.2. Bilans énergétiques annuels et mensuels

La *figure 4* représente les *productivités solaires mensuelles $PS_M$* produites en 1992, 1993 et 1994. La productivité solaire est plus élevée en mars - avril (entre 25 et 40 kWh.m$^{-2}$.mois$^{-1}$) comme on le prévoit car les besoins de chauffage restent importants alors que le rayonnement solaire est quasi perpendiculaire au plan des capteurs; elle reste comprise entre 15 et 25 kWh.m$^{-2}$.mois$^{-1}$ de mai à octobre et descend à des valeurs comprises entre 5 et 17 kWh.m$^{-2}$.mois$^{-1}$ de novembre à janvier. La production solaire est donc assez importante en hiver pour couvrir une part significative des besoins.

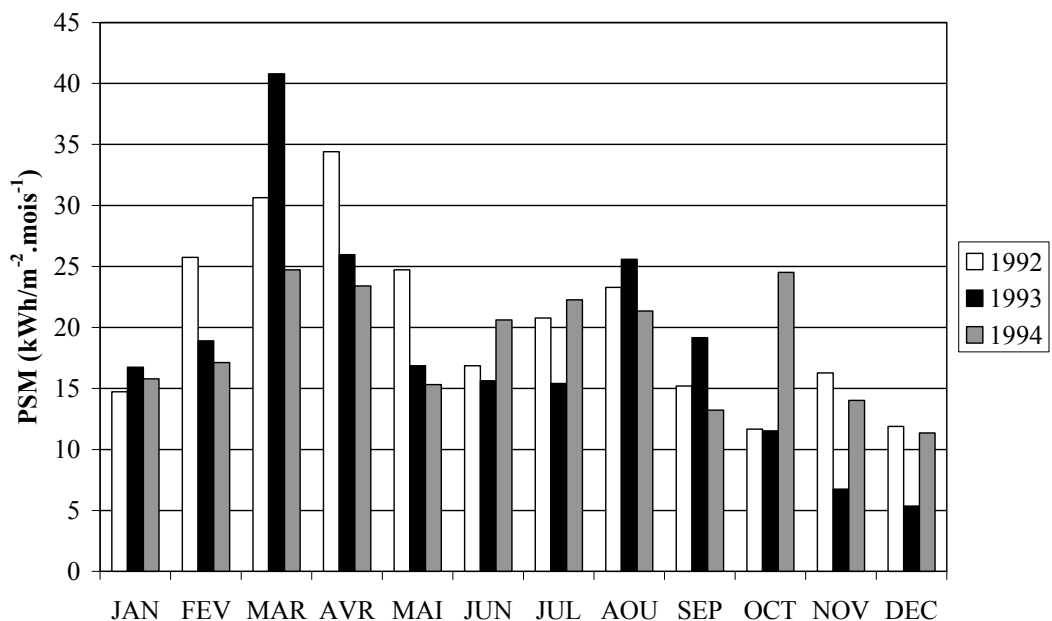

*Fig. 4. – Productivité solaire de l'installation*
Fig.4. – Solar productivity of solar plant

On peut comparer utilement les productivités solaires annuelles qui en résultent aux heures d'insolation mesurées à la station météorologique de Mulhouse et aux heures de fonctionnement effectif de l'installation solaire *(tableau II)*.

On remarque une assez bonne corrélation entre la durée de l'insolation mesurée à la station la plus proche et la durée de fonctionnement de l'installation pour les mois de chauffe de novembre à mars, l'écart se creusant pour les autres mois ou la production d'E.C.S. devient importante.

Les trois années ayant été nettement moins ensoleillées qu'en moyenne, il convient de corriger les productivités solaires mensuelles et annuelles en tenant compte de la durée d'insolation, mais également des températures extérieures afin de connaître leurs valeurs moyennes: *la productivité solaire annuelle pour une année moyenne est comprise entre* 267 *et* 293 kWh.m$^{-2}$.an$^{-1}$. Cette fourchette correspond aux résultats de calcul les plus significatifs obtenus par différentes régressions linéaires multiples effectuées sur les durées d'insolation mensuelles de Mulhouse et/ou Colmar, les températures extérieures moyennes mensuelles de Mulhouse et éventuellement les températures moyennes de stock mensuelles observées sur les trois années; les régressions effectuées par saisons-types de fonctionnement (P.S.D. continu ; P.S.D intermittent ; E.C.S. seule) donnent également d'assez bonnes corrélations, et les régressions utilisant simultanément les deux stations sont généralement les meilleures. Ces valeurs sont très proches des valeurs obtenues dans le dimensionnement par méthode E.S.I.M.; cependant l'étude mensuelle montre que *la méthode E.S.I.M. sous-estime la productivité solaire d'hiver (novembre à février inclus) du fait de la réduction du chauffage d'appoint permis par les apports passifs, ainsi que celle d'été (juin à août) par sous-estimation des consommations d'E.C.S.; elle surestime celle d'intersaisons car elle ne prend pas en compte de la production d'ECS dans cette période où les besoins en chauffage pour une maison bioclimatique sont plus réduits.* Les couplages avec l'enveloppe du bâtiment et la régulation du chauffage d'appoint par les apports solaires passifs et actifs ne sont donc pas correctement évalués par la méthode E.S.I.M. qui est une méthode globale mensuelle.

| | TABLEAU II / TABLE II *Productivités solaires mensuelles de l'installation* Monthly solar productivity of the plant | | | | | | | | | | | |
|---|---|---|---|---|---|---|---|---|---|---|---|---|
| | 1992 | | | | 1993 | | | | 1994 | | | |
| Mois | $HS_M$ | $D_M$ | $T_E$ | $PS_M$ | $HS_M$ | $D_M$ | $T_E$ | $PS_M$ | $HS_M$ | $D_M$ | $T_E$ | $PS_M$ |
| *JAN* | 78,5 | 61,5 | 0,7 | 14,7 | 73,9 | 65,6 | 4,6 | 16,8 | 68,8 | 63,5 | 3,7 | 15,8 |
| *FEV* | 111,8 | 86,7 | 3,5 | 25,7 | 91,8 | 65,4 | 1,3 | 18,9 | 49,0 | 71,5 | 3,6 | 17,1 |
| *MAR* | 103,3 | 104,2 | 7,2 | 30,6 | 177,6 | 134,0 | 6,1 | 40,8 | 104,0 | 102,6 | 10,3 | 24,7 |
| *AVR* | 157,8 | 115,4 | 9,9 | 34,4 | 177,0 | 107,0 | 12,3 | 26,0 | 114,6 | 110,2 | 9,1 | 23,4 |
| *MAI* | 214,6 | 123,9 | 16,1 | 24,7 | 179,3 | 76,8 | 15,7 | 16,9 | 152,5 | 70,7 | 14,6 | 15,3 |
| *JUN* | 150,9 | 83,9 | 18,0 | 16,9 | 200,3 | 72,6 | 18,0 | 15,6 | 212,6 | 114,3 | 18,6 | 20,6 |
| *JUL* | 214,3 | 124,2 | 20,8 | 20,8 | 209,5 | 70,7 | 18,8 | 15,4 | 265,0 | 124,3 | 22,9 | 22,3 |
| *AOU* | 222,0 | 135,1 | 21,9 | 23,3 | 255,6 | 115,5 | 19,1 | 25,6 | 237,0 | 108,0 | 20,8 | 21,4 |
| *SEP* | 168,0 | 96,4 | 16,1 | 15,2 | 120,2 | 82,3 | 14,6 | 19,2 | 95,6 | 79,0 | 15,3 | 13,2 |
| *OCT* | 47,6 | 38,8 | 8,6 | 11,7 | 42,9 | 40,0 | 9,2 | 11,5 | 130,5 | 105,0 | 10,3 | 24,5 |
| *NOV* | 51,7 | 51,6 | 7,6 | 16,3 | 31,1 | 28,8 | 1,9 | 6,7 | 67,0 | 65,0 | 9,1 | 14,0 |
| *DEC* | 58,6 | 50,1 | 2,6 | 11,9 | 27,4 | 21,3 | 5,5 | 5,4 | 46,5 | 48,6 | 4,8 | 11,3 |
| ANNÉE | 1579,1 | 1071,8 | 11,1 | 246,2 | 1586,6 | 880,0 | 10,7 | 218,7 | 1543,1 | 1062,7 | 12,0 | 223,7 |

La *figure 5* permet de mieux comprendre l'importance du choix de l'inclinaison i des capteurs dans la gestion de la ressource solaire. La *productivité solaire horaire* moyenne mensuelle $PSH_M$ présente une homogénéité remarquable autour de 230 W.m$^{-2}$, avec une plage une peu plus élevée autour de 280 W.m$^{-2}$ en intersaison et un peu plus faible vers 200W.m$^{-2}$ en période estivale. Cette régularité est liée au choix d'optimisation du fonctionnement par l'inclinaison à 58° pour une latitude de 47,6° permettant une très bonne productivité en hiver, une autonomie maximale de chauffage en intersaison et une autonomie totale de production d'E.C.S. en été sans beaucoup d'excédent.

Les *figures 6 et 7* confirment ce choix de manière évidente, l'autonomie énergétique totale (chauffage + E.C.S.) étant assurée de mai à septembre compris. Les relevés quotidiens révèlent une autonomie totale pendant 179 à 213 jours par an (soit 49% à 58,3% de l'année) et une saison de chauffe effective réduite entre 92 et 118 jours par an (soit 25,2% à 32,3% de l'année), dans une région où la saison de chauffe débute en septembre et termine en mai, voire en juin. La *figure 8* montre que le *taux de couverture solaire annuel corrigé* $\tau_{AC}$ fluctue selon les conditions tre 38,2% et 50,3% en 1993 et 1994; ces taux sont excellents, malgré des températures extérieures sur site systématiquement plus faibles qu'à la station de Mulhouse en hiver et intersaison, un ensoleillement inférieur à la moyenne et une consommation d'énergie prenant en compte tous les besoins domestiques hormis la cuisine. Ils seraient de l'ordre de 45,6% à 47,9% pour une année standard. Seul un recul sur une dizaine d'année permettra de donner une estimation statistique fiable du taux de couverture solaire corrigé annuel.

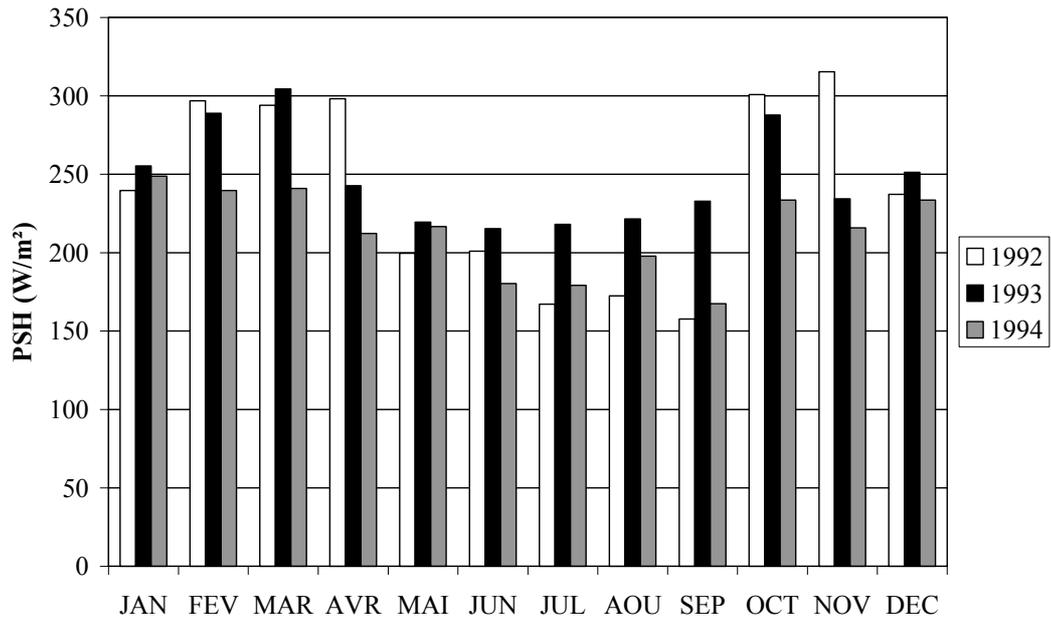

*Fig. 5. – Productivité solaire horaire de l'installation*
Fig. 5. – Per-hour solar productivity of solar plant

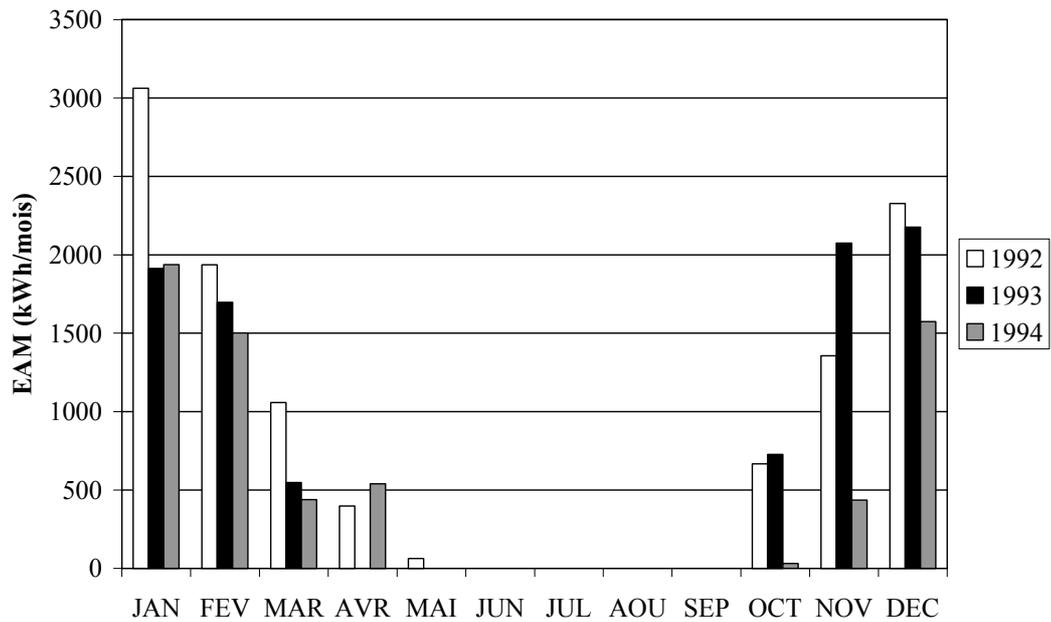

*Fig. 6. – Énergie d'appoint totale mensuelle*
Fig. 6. – Monthly total supply energy consumption

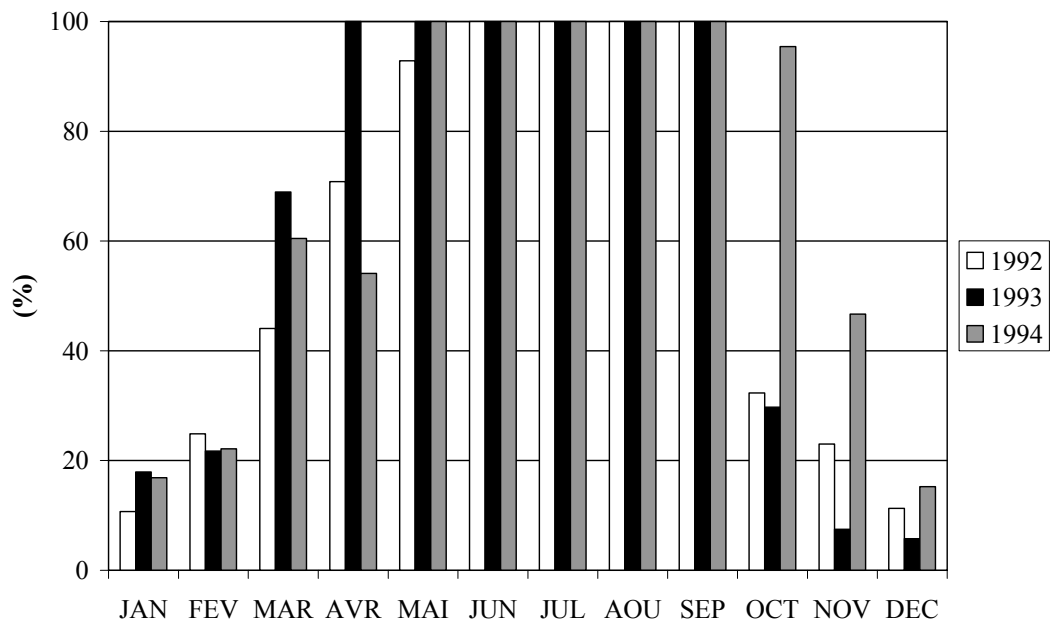

*Fig. 7. – Taux de couverture solaire mensuelle corrigé de l'installation*
Fig. 7. – Monthly corrected solar covering ration of solar plant

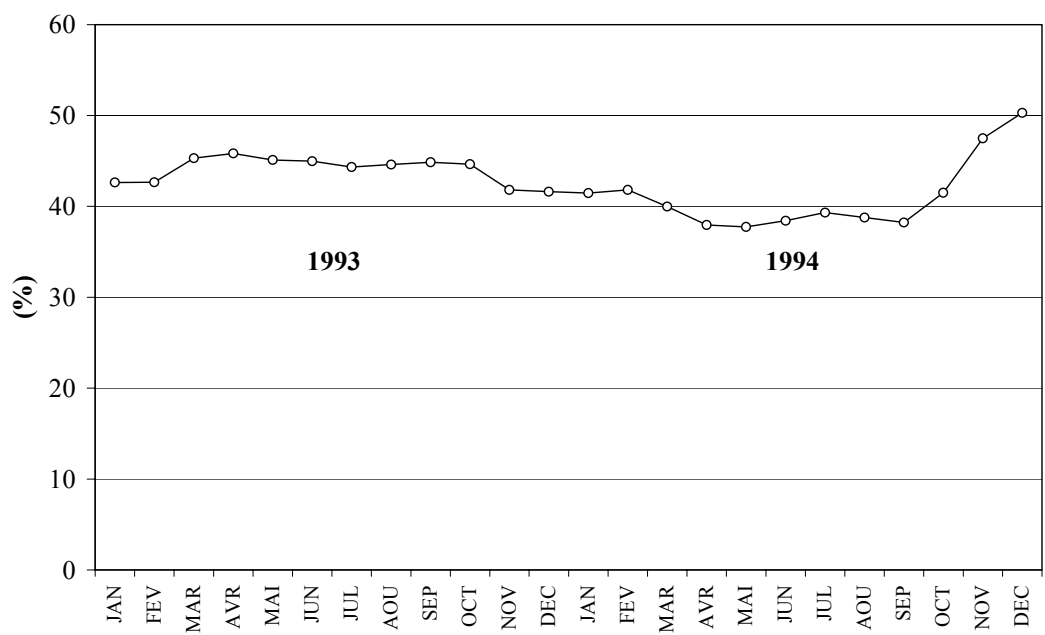

*Fig. 8. – Taux de couverture solaire annuel corrigé*
Fig. 8. – Annual corrected solar covering ration of solar plant

L'économie d'énergie globale par rapport à une habitation classique est en réalité plus élevée si l'on tient compte des *apports solaires passifs* gratuits qui représentent ici 69% de l'énergie solaire directe produite par les capteurs en hiver et intersaison: les apports passifs, évalués par la méthode C.S.T.B. [15], sont de l'ordre de 3500 kWh.an$^{-1}$ pour une année moyenne; ils ont représenté entre 14% et 17% des besoins de l'habitation de 1992 à 1994, l'énergie économisée par l'installation solaire active couvrant alors 36% à 42% de ces besoins, ce qui donne pour l'ensemble des apports solaires {actifs + passifs} un taux de couverture solaire total de 48% à 59%.

De fait, une habitation de mêmes surface et volume habitables, d'isolation réglementaire standard ($G = 0,9$ W.m$^{-3}$.K$^{-1}$) consommerait en conditions moyennes environ 24000 kWh par an, hors rendements et E.C.S. (méthode C.S.T.B.), soit plus de 2,5 fois l'énergie consommée ici (plus de 4,4 fois en tenant compte des rendements de chaudière).

## 5. ANALYSE DU FONCTIONNEMENT DU P.S.D. MIXTE

Le bilan énergétique réel très satisfaisant du système solaire à P.S.D. mixte établi précédemment peut être expliqué et analysé par une étude plus fine des relevés et des enregistrements automatiques quotidiens effectués sur l'installation, sur des séquences caractéristiques. Les prévisions de comportement de l'installation reposent sur *l'effet d'asservissement de l'appoint* par les deux informations physiques que sont *la température globale de la dalle* –qui contient elle-même l'information thermique sur la séquence solaire couvrant les trois journées précédentes– et *la température intérieure de l'habitation*, représentative de l'ensoleillement du moment grâce à la conception bioclimatique de l'habitation (grands apports solaires passifs, faible inertie des parois internes) . C'est l'ensemble de ces paramètres de couplage qui assure la bonne performance du système; on peut ainsi représenter le fonctionnement du système par le schéma de principe de la *figure 9*.

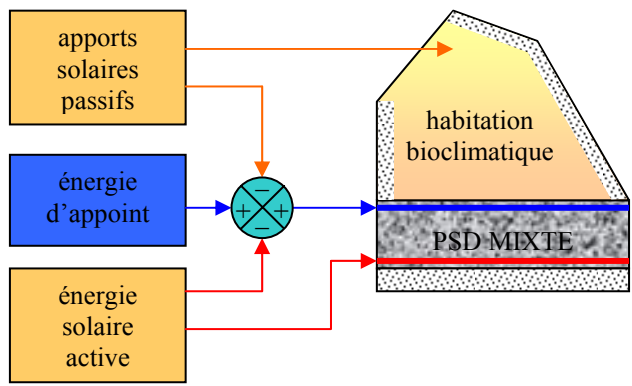

*Fig. 9. – Couplage solaire-appoint-enveloppe dans le PSD mixte*
Fig. 9. – Coupling between solar energy-supply energy-habitation in mixed direct solar floor

### 5.1. Chauffage solaire + chauffage d'appoint en période hivernale très froide

La séquence du 18 au 22 février 1992 fait suite à plusieurs jours couverts très froids *(figures 10, 11, 12, 13)*. Au début de la première belle journée de la séquence, le 19 février, pour maintenir une température ambiante minimale de 19°C après une température extérieure nocturne entre -5°C et -12 °C, le circuit d'appoint en surface de dalle se referme totalement 50 minutes après le démarrage du chauffage solaire, et ne se rouvre que légèrement le lendemain, pour se refermer entièrement moins de 15 minutes après l'enclenchement du circuit solaire, grâce à l'effet de stockage solaire dans la dalle. Le lendemain, il ne se rouvre que très peu pendant environ 2,6 heures, et pas du tout le surlendemain, l'apport solaire étant suffisant malgré une température extérieure toujours négative. Le temps de transit $\Delta t_{SA}$ enregistré est de l'ordre de 4,1 h $\pm$ 0,2 h (l'inertie thermique des tubes fortement isolés en chaufferie explique l'apparente durée de fonctionnement du P.S.D. plus longue que la durée d'ensoleillement; ce phénomène n'existe pas pour le réseau d'appoint dont le circulateur est constamment en fonction).

L'étude des rendements solaires journaliers et de la productivité solaire horaire *(tableau III)* montre que l'influence du chauffage d'appoint est négligeable sur ces paramètres et non corrélable, contrairement à la température extérieure qui joue un rôle dans le temps et l'énergie nécessaires à la mise en température des capteurs. Ainsi les deux journées les plus froides du 19 et du 20 février, autour de -5°C en moyenne jour-nuit montrent un même rendement de 45% et une productivité solaire horaire identique à 306 W.m$^{-2}$, alors que l'appoint a fourni près de trois fois moins d'énergie à la surface de la dalle. Les deux jours suivants voient le rendement s'élever à 49,4% du fait de l'élévation de température, alors que l'appoint gaz a été à peu près identique du 20 au 21. Enfin, la journée du 24 montre un rendement et une PSH nettement supérieurs de 54,2% et 315 W.m$^{-2}$ respectivement pour une quantité d'énergie d'appoint fournie à la dalle similaire à celle du 22 car la température extérieure moyenne est passée à 1,9°C. Le cas des journées peu ensoleillées du 18 et du 23 février montre que la productivité solaire horaire est restée élevée (et même meilleure le 18 alors que l'appoint était important) et le rendement identique mais plus faible, ce dernier étant sensible à la valeur de l'irradiation solaire. Les énergies enregistrées sont données à 1 kWh près.

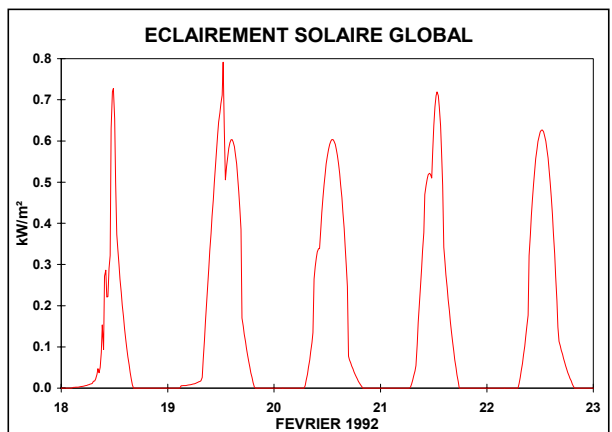

*Fig. 10. – Éclairement solaire global E dans le plan des capteurs (du 18 au 22 février 1992)*
Fig. 10. – Global solar flow *E* received on solar collectors plan (1992, 18 to 22 February)

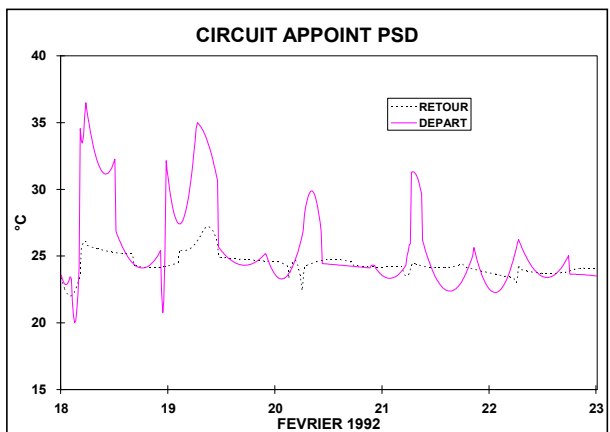

*Fig. 12. – Températures du circuit d'appoint en surface de dalle (du 18 au 22 février 1992)*
Fig. 12. – Supply energy heating tube network temperatures (1992, 18 to 22 February)

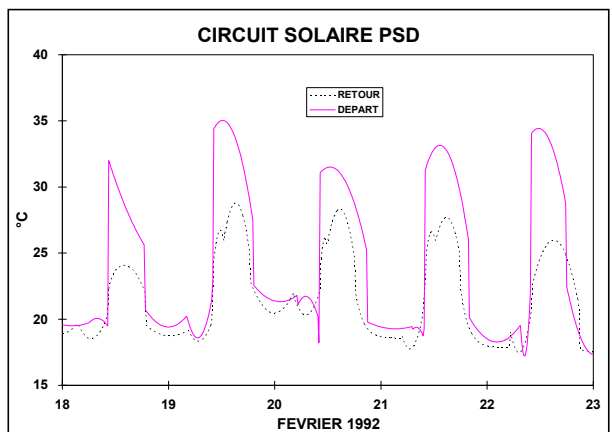

*Fig. 11. – Températures du circuit solaire dans la dalle (du 18 au 22 février 1992)*
Fig. 11. – Solar heating tube network temperatures in floor (1992, 18 to 22 February)

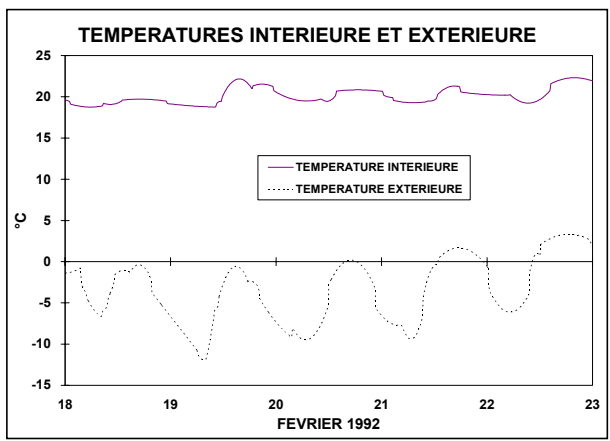

*Fig. 13. – Températures intérieure et extérieure (du 18 au 22 février 1992)*
Fig. 13. – Inner and outer temperatures (1992, 18 to 22 February)

| TABLEAU III – TABLE III ||||||||||
|---|---|---|---|---|---|---|---|---|
| *Séquence chauffage solaire + appoint en période hivernale très froide* ||||||||||
| Solar + complementary heating during very cold winter period ||||||||||
| FEVRIER 1992 | 18 | 19 | 20 | 21 | 22 | 23 | 24 | *MOIS* |
| $ES_0$ (kWh) | 33,7 | 77,5 | 64,8 | 59,7 | 61,9 | 37,4 | 70,6 | *1028,5* |
| $ES$ (kWh) | 13,0 | 34,8 | 29,4 | 29,4 | 30,7 | 12,3 | 38,2 | *437,5* |
| $D$ (h) | 2,44 | 6,66 | 5,66 | 5,60 | 6,05 | 2,48 | 7,14 | *86,7* |
| $T_E$ (°C) | -3,1 | -5,8 | -4,2 | -2,9 | -0,53 | 4,1 | 1,9 | *2,5* |
| $T_I$ (°C) | 19,3 | 20,2 | 20,2 | 20,1 | 20,8 | 19,9 | 20,2 | *20,0* |
| $EA_D$ (kWh) | 44 | 40 | 15 | 12 | 8 | 0 | 7 | *651* |
| η (%) | 38,5 | 44,9 | 45,3 | 49,2 | 49,6 | 32,8 | 54,2 | *42,5* |
| $PSH$ (W.m$^{-2}$) | 313 | 307 | 305 | 309 | 299 | 291 | 315 | *296,8* |

## 5.2. Chauffage solaire sans aucun appoint en période hivernale très froide

Une séquence quasi identique s'est déroulée du 23 au 27 février 1993 avec des températures extérieures similaires entre 0°C et -12°C *(figures 14, 15, 16)* mais le chauffage d'appoint a été coupé depuis le 22 février et remis en fonction le 2 mars. On constate que l'énergie solaire suffit à maintenir la température intérieure entre 17 à 18°C la nuit et 20 à 21°C le jour, et il est intéressant de comparer les paramètres de l'installation solaire durant cette période à celles de la séquence précédente avec appoint. Les résultats sont regroupés dans le *tableau IV*.

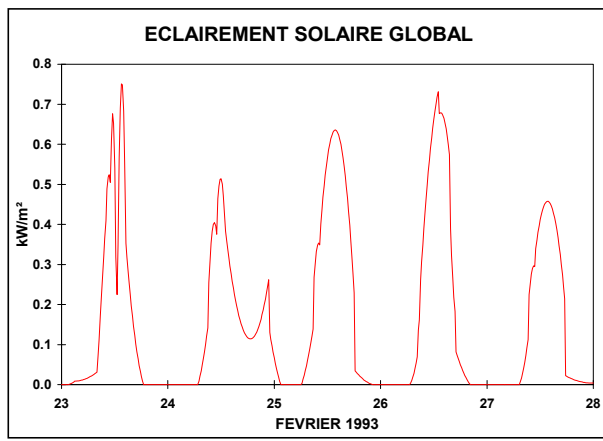

*Fig. 14. – Éclairement solaire global E dans le plan des capteurs (du 23 au 27 février 1993)*
Fig. 14. – Global solar flow *E* received on solar collectors plan (1993, 23 to 27 February)

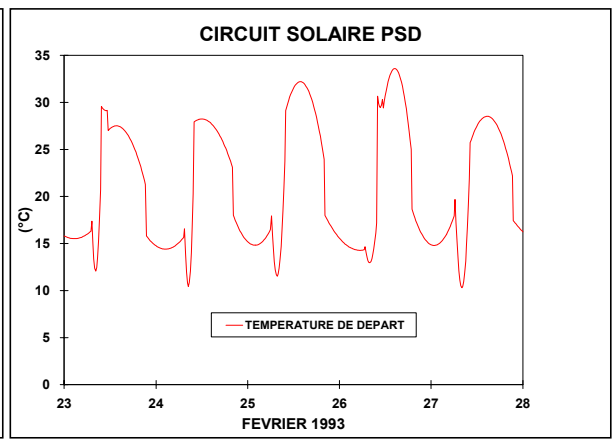

*Fig. 15. – Température du circuit solaire dans la dalle (du 23 au 27 février 1993)*
Fig. 15. – Solar heating tube network temperature in floor (1993, 23 to 27 February)

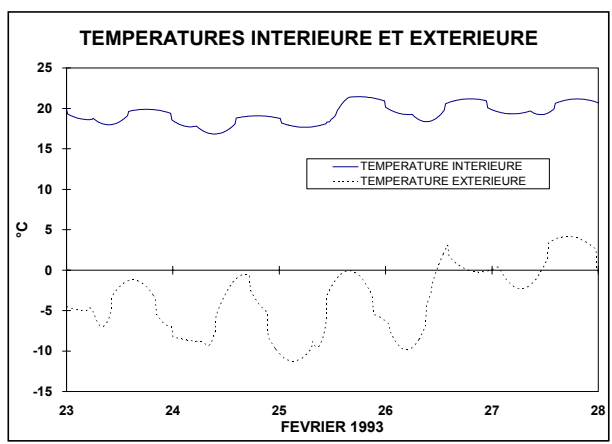

*Fig. 16. – Températures intérieure et extérieure (du 23 au 27 février 1993)*
Fig. 16. – Inner and outer temperatures (1993, 23 to 27 February)

On constate que la productivité solaire horaire varie dans une fourchette identique (compte tenu des incertitudes de mesure d'énergie à 1 kWh près) à celle de la séquence de février 1992 ; elle est plus sensible aux passages nuageux et au rapport éclairement diffus/éclairement direct, alors que le rendement solaire est plus sensible à la température extérieure. Ainsi les journées du 26 février 1993 et du 21 février 1992, toutes deux autour de -3°C, donnent un rendement d'installation solaire d'environ 48,5% ; en revanche, la productivité solaire horaire est de 363 W.m$^{-2}$ pour la première journée et de 309 W.m$^{-2}$ pour la seconde du fait d'un éclairement solaire $ES_0$ nettement plus élevé dans le premier cas. De même les journées du 24 février 1993 et du 19 février 1992, de même température moyenne, donnent le même rendement de 45,5% et une productivité *PSH* moins grande dans le premier cas du fait d'un moindre éclairement solaire. Si l'on compare enfin deux journées quasi identiques comme celles du 23 février 1993 et du 20 février 1992 ($T_E \approx$ -4,2°C ; $ES_0 \approx$ 62 kWh), on observe une productivité solaire horaire un peu moins bonne pour la première mais un rendement un peu supérieur. L'explication est à rechercher dans la répartition journalière du flux solaire lors de ces deux journées, l'irradiation solaire instantanée ayant atteint 760W.m$^{-2}$ le 23 février 1993 et seulement 610 W.m$^{-2}$ le 20 février 1992 (*voir figures 10 et 14*). On peut conclure de l'examen de ces deux séquences que *la productivité solaire du P.S.D. mixte est apparemment insensible au chauffage d'appoint en surface de dalle; il en est de même du rendement solaire, à température extérieure égale.*

| TABLEAU IV – TABLE IV *Séquence chauffage solaire sans appoint en période hivernale très froide* Solar heating without complementary heating during very cold winter period | | | | | | |
|---|---|---|---|---|---|---|
| Février 1993 | 23 | 24 | 25 | 26 | 27 | *Mois* |
| $ES_0$ (kWh) | 60,0 | 63,3 | 80,1 | 77,3 | 57,1 | *846,8* |
| $ES$ (kWh) | 30,9 | 29,1 | 40,9 | 37,3 | 29,1 | *321,2* |
| $D$ (h) | 6,20 | 6,40 | 7,20 | 6,05 | 6,29 | *65,4* |
| $T_E$ (°C) | -4,2 | -5,8 | -5,7 | -3,2 | 1,2 | *0,37* |
| $T_I$ (°C) | 19,0 | 18,1 | 19,4 | 19,9 | 20,1 | *19,0* |
| $EA_D$ (kWh) | 0 | 0 | 0 | 0 | 0 | _ |
| $\eta$ (%) | 51,6 | 46,0 | 51,1 | 48,2 | 51,0 | *37,9* |
| $PSH$ (W.m$^{-2}$) | 293 | 268 | 335 | 363 | 272 | *288,9* |

### 5.3. Chauffage solaire seul en intersaison, avec effet de stockage

L'inertie du P.S.D. est particulièrement importante en intersaison: combinée aux apports passifs de l'habitation, elle permet de se passer totalement de chauffage d'appoint lors de périodes comportant deux à trois jours sans soleil. On peut à cet égard examiner les enregistrements effectués du 5 au 10 novembre 1992, séquence caractérisée par une alternance de belles journées et de jours sans soleil par une température de 7,5°C en moyenne *(figures 17, 18, 19)*. Là encore, l'énergie solaire suffit à maintenir la température ambiante autour de 20°C. Les performances de l'installation sont résumées dans le *tableau V*. On remarquera la valeur plus grande du rendement et de la productivité solaire horaire en intersaison du fait de la basse température du PSD et de la température extérieure encore clémente.

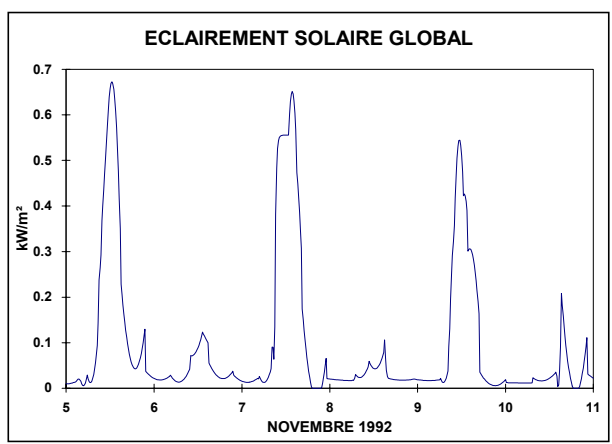

*Fig. 17. – Éclairement solaire global E dans le plan des capteurs (du 5 au 10 novembre 1992)*
Fig. 17. – Global solar flow E received on solar collectors plan (1992, 5 to 10 November)

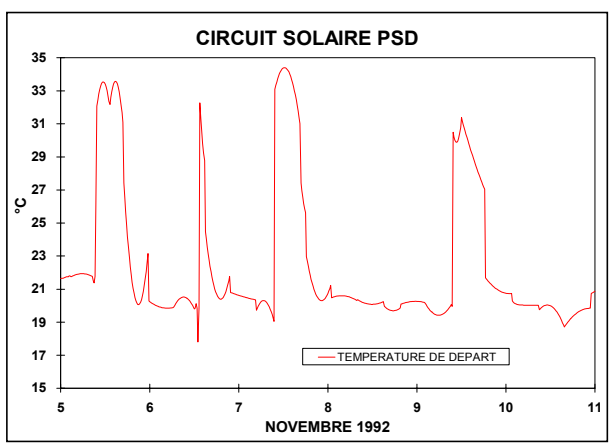

*Fig. 18. – Température du circuit solaire dans la dalle (du 5 au 10 novembre 1992)*
Fig. 18. – Solar heating tube network temperature in floor (1992, 5 to 10 November)

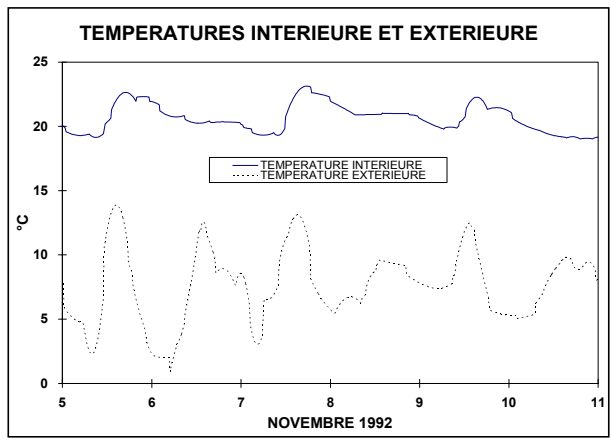

*Fig. 19. – Températures intérieure et extérieure
(du 5 au 10 novembre 1992)*
Fig. 19. – Inner and outer temperatures
(1992, 5 to 10 November)

| TABLEAU V – TABLE V<br>*Séquence chauffage solaire seul en intersaison, avec effet de stockage*<br>Solar heating alone during interseason, with storage effect | | | | | | | |
|---|---|---|---|---|---|---|---|
| Novembre 1992 | 5 | 6 | 7 | 8 | 9 | 10 | *Mois* |
| $ES_0$ (kWh) | 67,6 | 10,9 | 70,92 | 6,4 | 50,63 | 6,7 | *580,0* |
| $ES$ (kWh) | 38,2 | 0 | 37,31 | 0 | 24,57 | 0 | *276,6* |
| $D$ (h) | 6,75 | 0,05 | 6,24 | 0,00 | 5,06 | 0,03 | *51,59* |
| $T_E$ (°C) | 7,0 | 6,7 | 8,1 | 7,8 | 8,1 | 7,6 | *7,3* |
| $T_I$ (°C) | 20,7 | 20,6 | 21,0 | 21,0 | 20,9 | 20,9 | *19,6* |
| $EA_D$ (kWh) | 0 | 0 | 0 | 0 | 0 | 0 | _ |
| η (%) | 56,6 | 0,0 | 52,6 | 0,0 | 48,5 | 0,0 | *47,7* |
| $PSH$ (W.m$^{-2}$) | 333 | 0 | 352 | 0 | 286 | 0 | *315,3* |

### 5.4. Chauffage et E.C.S. solaires seuls sans appoint en intersaison

Le fonctionnement en intersaison où l'autonomie totale ou quasi totale en chauffage et E.C.S. est atteinte peut être étudié lors de la séquence du 7 au 12 avril 1992 *(figures 20, 21, 22, 23)* lorsque la température de consigne de surchauffe de dalle $T_{CD}$ est dépassée, le circuit solaire du P.S.D. s'arrêtant alors pour laisser le circuit solaire E.C.S. absorber seul l'énergie solaire. Les résultats sont rassemblés dans le *tableau VI*.

On observe la baisse de rendement entraînée par l'arrêt du chauffage solaire, qui passe d'environ 56% lorsque les deux circuits solaires fonctionnent en parallèle, à 24% lorsque seul le circuit solaire E.C.S. est en fonction. La productivité solaire horaire est maximale en cette période (chauffage solaire + stock E.C.S. froid), et peut être élevée même avec un mauvais rendement global par journée faiblement ensoleillée comme celle du 13 avril.

Les mesures manuelles de température de surface de dalle ont montré qu'en aucun cas, elle n'a dépassé 25°C, que ce soit dans cette période charnière de forts apports passifs ou dans la période hivernale. Les pointes de 24 à 25°C relevées les 11 et 12 avril sont dues aux apports passifs incontrôlés en l'absence des propriétaires.

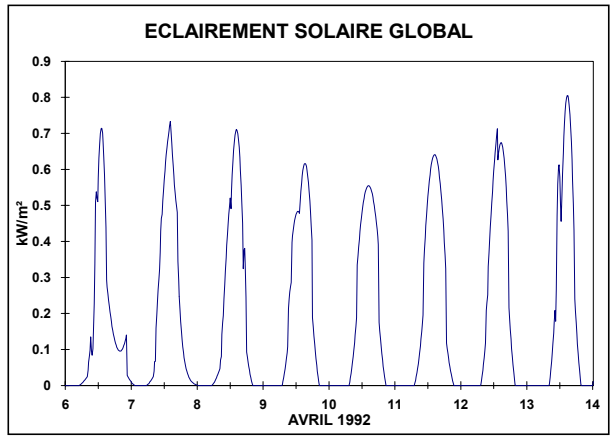

Fig. 20. – *Éclairement solaire global E dans le plan des capteurs (du 6 au 13 avril 1992)*
Fig. 20. – Global solar flow *E* received on solar collectors plan (1992, 6 to 13 April)

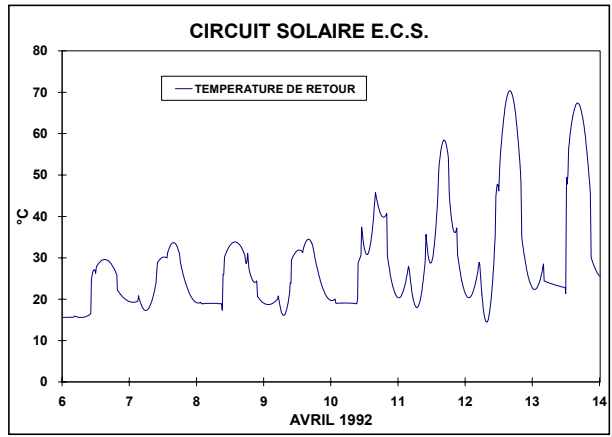

Fig. 22. – *Température du circuit solaire d'ECS (du 6 au 13 avril 1992)*
Fig. 22. – SHW solar tube temperature (1992, 6 to 13 April)

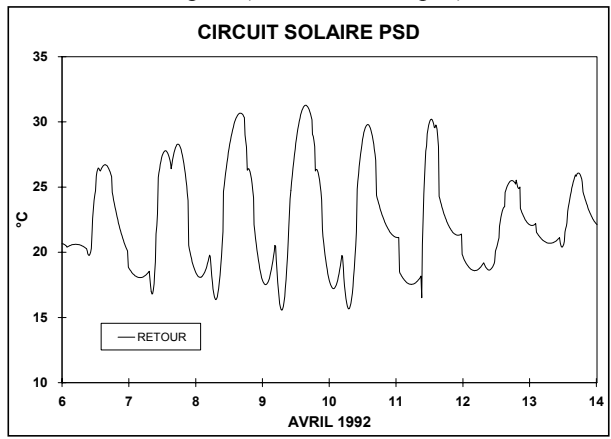

Fig. 21. – *Température du circuit solaire dans la dalle (du 6 au 13 avril 1992)*
Fig. 21. – Solar heating tube network temperature in floor (1992, 6 to 13 April)

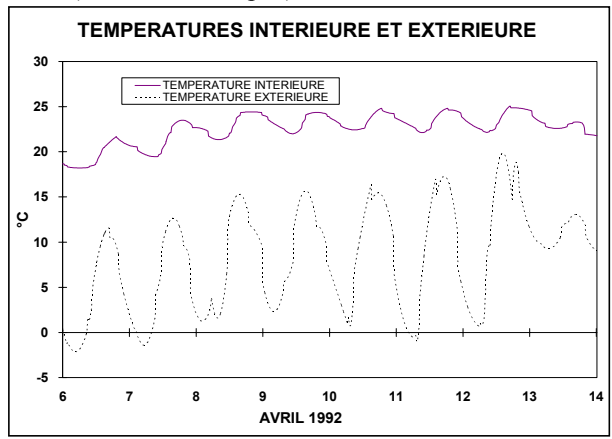

Fig. 23. – *Températures intérieure et extérieure (du 6 au 13 avril 1992)*
Fig. 23. – Inner and outer temperatures (1992, 6 to 13 April)

| TABLEAU VI – TABLE VI *Séquence chauffage et E.C.S. solaires autonomes en intersaison* Autonomic solar heating and sanitary warm water production during interseason ||||||||||
|---|---|---|---|---|---|---|---|---|---|
| Avril 1992 | 6 | 7 | 8 | 9 | 10 | 11 | 12 | 13 | *MOIS* |
| $ES_0$ (kWh) | 66,97 | 81,43 | 77,13 | 76,90 | 74,12 | 81,72 | 82,27 | 81,77 | *1565,1* |
| $ES$ (kWh) | 37,54 | 47,09 | 43,68 | 39,58 | 28,66 | 30,03 | 19,79 | 9,55 | *584,9* |
| $D$ (h) | 7,43 | 8,42 | 7,76 | 7,51 | 4,41 | 5,04 | 5,25 | 1,80 | *115,4* |
| $T_E$ (°C) | 4,0 | 5,4 | 8,0 | 8,9 | 9,3 | 8,7 | 11,0 | 10,8 | *9,8* |
| $T_I$ (°C) | 19,5 | 21,3 | 22,9 | 23,3 | 23,4 | 23,5 | 23,6 | 22,8 | *21,6* |
| $EA$ (kWh) | 57,8 | 8,3 | 6,2 | 0 | 0 | 0 | 0 | 0 | *398,7* |
| η (%) | 56,1 | 57,8 | 56,6 | 51,5 | 38,7 | 36,7 | 24,1 | 11,67 | *37,4* |
| $PSH$ (W.m$^{-2}$) | 297 | 329 | 331 | 310 | 382 | 350 | 222 | 312 | *297,7* |

Les relevés des années suivantes montrent une meilleure gestion de ces apports gratuits en avril-mai, et les risques de surchauffe en intersaison sont effectivement nuls, l'aquastat de coupure jouant parfaitement son rôle. Une température moyenne inférieure à 23°C est maintenue sans problèmes.

Le choix de l'épaisseur de la dalle a également des conséquences pour la tenue thermique de l'habitation en été; pendant cette saison, les enregistrements effectués (non donnés ici) montrent que l'habitation reste fraîche (entre 20 et 25°C) en plein été grâce à l'inertie thermique de la dalle sur cave et à la très bonne isolation de l'enveloppe, à condition que les surfaces vitrées soient correctement occultées au sud, le rafraîchissement naturel de nuit par les ouvrants suffisant à abaisser la température et la dalle restant suffisamment froide même en journée.

## 6. CONCLUSION

La réalisation et le suivi d'une installation de chauffage solaire par Plancher Solaire Direct Mixte épais dans une habitation bioclimatique en région à climat continental de faible ensoleillement a permis de démontrer la bonne complémentarité du chauffage solaire actif et du chauffage solaire passif.

Le P.S.D. mixte assure la double fonction de stockage solaire par inertie thermique, complémentaire de la faible inertie de l'enveloppe de l'habitat, et de régulation de l'énergie d'appoint en surface de P.S.D. par asservissement thermique grâce à la fois aux apports passifs, à la faible inertie thermique de l'enveloppe de l'habitation et à l'épaisseur de la dalle.

Les mécanismes physiques de cette régulation expérimentalement observée font intervenir de nombreux paramètres physiques, climatiques et architecturaux, caractérisés par des couplages élevés dont il reste à simuler l'étendue. Cependant, l'étude énergétique du système a montré le très bon comportement de l'installation solaire dont le rendement et la productivité ne semblent pas affectés par l'utilisation conjointe de la dalle et du stock d'eau chaude sanitaire par les circuits d'appoint, au vu des mesures et des essais effectuées sur les trois années de 1992 à 1994.

Les logiciels couramment utilisés en ingénierie solaire, s'ils permettent une évaluation globale des performances des P.S.D. en moyenne annuelle, ne rendent cependant pas compte du comportement dynamique d'une telle installation et des couplages solaire-appoint en interaction avec l'enveloppe. Des modèles bidimensionnels de dalle à double nappe, associés à des modèles numériques zonaux plus fins de type modulaire pour l'enveloppe doivent encore être développés pour simuler les interactions entre circuits solaires et circuits d'appoint, tant en chauffage qu'en production d'E.C.S.; vérifier que l'apport d'énergie d'appoint en surface de dalle solaire n'obère pas les performances solaires; affiner les choix de dimensionnement présentés plus haut; et approcher les productivités journalières, mensuelles et annuelles avec une précision satisfaisante.

L'emploi des critères de productivité solaire horaire, d'énergie économisée et de taux de couverture solaire corrigé offre l'avantage de pouvoir comparer les installations solaires en fonctionnement réel, leurs performances intrinsèques et les performances globales des systèmes couplés {installation solaire / appoint / enveloppe du bâtiment}, ainsi que leur évolution sur plusieurs années.

La technique de chauffage solaire basse température par P.S.D. mixte avec appoint indépendant en surface, de grandes fiabilité et simplicité, doit donc permettre une meilleure diffusion de l'utilisation de l'énergie solaire à tous les types d'habitation individuelle, y compris les bâtiments à hautes performances énergétiques caractérisées par une très grande isolation thermique, une faible inertie intérieure des parois et de grandes ouvertures vitrées du côté ensoleillé. Elle doit permettre également une meilleure prise en compte du chauffage solaire actif dans les règles d'urbanisme, les réglementations à venir et l'architecture de demain.

A cet égard, elle complète utilement la technique du P.S.D. mince à appoint intégré particulièrement adaptée aux bâtiments collectifs et aux habitations individuelles classiques.

## BIBLIOGRAPHIE

---

## ABRIDGED ENGLISH VERSION
## *DOUBLE DIRECT SOLAR FLOOR HEATING IN BIOCLIMATIC HABITATION*
### *Design and real thermal balance*

The Direct Solar Floor Heating Technique (Plancher Solaire Direct in french) was invented several years ago by the E.S.I.M. (the Engineering High School of Marseille) [1] and is largely employed in heating systems using solar energy in France, essentially in domestic habitations [4][5][6][7][8]. This technique is simpler and more economic than the previous ones [2] since there are no heat exchangers between solar collectors and heat storage equipment, and because the heating floor acts as a heat emitter as well as a storage device by means of its concrete thickness (approximately 30 cm). However an important disadvantage still remains since another heating installation providing complementary energy and full heating power during sunless periods, is necessary.

Several authors [9][10][11] made an attempt in reducing the investment cost by reducing the thickness of the floor by a factor of two and by using the same heating tubes for both solar and complementary energies. They used a computer controlled technology in order to regulate the complete system and tested it in an experimental plant. By numerical simulation they decuced that it is a convenient way to produce solar energy even if this reduction leads to a light decrease of solar productivity. They pointed out that this Integrated Supply-Direct Solar Floor Heating System is efficient only if the thermal inertia of the house walls is sufficiently high and if the passive solar heating capability of the house is small enough.

We studied another way of improving the Solar Direct Floor Heating technique consisting in a thick concrete floor that is very well insulated on the underside, with two heating tubes networks embedded in it *(fig. 2)*: the solar one at the bottom of the floor in order to warm its whole mass; the second one – which is heated by complementary energy (gas)– a few centimeters under the upper surface of the floor. This disposition allows a thermal regulation of the latter by the former. We showed that this new technique is very efficient when coupled with a bioclimatic architecture – that is to say in a well oriented and insulated building with low thermal inertia walls and great passive solar energy flow through the windows *(fig. 9)*.

We have implemented this Mixed Direct Solar Floor Heating Technique (P.S.D Mixte in french) in a bioclimatic house that was built in Alsace (France) which is a cold and not very sunny region *(fig. 1, table I)*. The gas consumption, solar production, working time of solar pumps, sanitary hot water (E.C.S. in french) consumption were measured every day and several physical factors (such as solar energy flow, external and internal temperatures, solar floor and E.C.S. heating temperatures and pipe flow, complementary heating temperatures) were computer recorded, so that we are able to draw up a precise energy balance of this solar system over the last three years.

A short description of the solar and complementary energy systems can be given as follows *(fig. 3)*:

- heating floor surface and thickness: 89 m² and 26 cm respectively; habitable surface and volume of the house: 132 m² and 330 m³; sanitary hot water storage volume: 0.5 m³
- surface of solar collectors: 17 m²; optical coefficient: 0.68; thermal conductance: 4.2 W.m$^{-2}$.K$^{-1}$; incline: 58°; heating-floor solar tubes spacing: 20 cm
- complementary supply energy: natural gas; heating power of boiler: 18 kW; tubes spacing of the floor heating: 30 cm

We obtain a very good energy balance which can be described by the following results:

- the *annual corrected solar covering ratio* $\tau_{AC}$, which represents the ratio of the energy saved $EE_A$ to the total energy need ($EE_A+EA_A$), ranges from 40 to 55%; it depends only on meteorological conditions *(fig. 8)*;
- total energetic autonomy is obtained during 179 to 213 days a year (49% to 58.3% of a year); heating period is reduced to a very short time (92 to 118 days a year – 25.2 % to 32.3% of a year), compared to at least eight months, as usual in this region *(fig. 6 and 7)*;
- the *annual solar productivity* $PS_A$ ranges between 267 and 293 kWh.m$^{-2}$ *(fig. 4)*; the solar efficiency η of the installation is independent of the fact that complementary energy is supplied or not at heating floor surface, but depends on external temperature and solar energy production mode (heating-floor / E.C.S.) *(tables III-VI)*;
- the *per-hour solar productivity PSH* is a very representative factor of the solar system performance; it remains almost constant over characteristic periods in the year ranging between 180 and 300 W.m$^{-2}$ *(fig. 5)* and really independent of complementary energy supply;
- the examination of automatic records leads to a good understanding of the real functioning of the Mixed Direct Solar Floor Heating when coupled with bioclimatic architecture *(fig. 10-23)*.

## Conclusion

This new solar heating technique seems to be very efficient and can be applied at least to all types of well-oriented and well-insulated habitations. It allows the benefit of floor-heating comfort. Its simplicity and low cost implementation can lead to a rapid expansion in the field of domestic habitation and thus, to considerable energy saving.